\documentclass{scrartcl}

\usepackage[english]{babel}
\usepackage{booktabs}
\usepackage{bbm}
\usepackage{amsfonts}
\usepackage{natbib}
\usepackage{comment}
\usepackage{booktabs}
\usepackage{amsmath}
\usepackage{graphicx}
\usepackage{bbm}
\usepackage{amsfonts}
\usepackage{algorithm2e}
\usepackage{natbib}

\usepackage{placeins}

\usepackage[colorlinks=true, allcolors=blue]{hyperref}

\usepackage[letterpaper,top=2cm,bottom=2cm,left=2cm,right=2cm,marginparwidth=1.75cm]{geometry}

\title{Variational Inference for Semiparametric Bayesian Novelty Detection in Large Datasets}
\author{Luca Benedetti, Eric Boniardi, Leonardo Chiani,\\
Jacopo Ghirri, Marta Mastropietro,
Andrea Cappozzo, \\
\textit{Politecnico di Milano}\\
Francesco Denti\footnote{francesco.denti@unicatt.it}\\
\textit{Università Cattolica del Sacro Cuore, Milan}\\}

\date{}

\begin{document}

\maketitle
\abstract{
After being trained on a fully-labeled training set, where the observations are grouped into a certain number of known classes, novelty detection methods aim to classify the instances of an unlabeled test set while allowing for the presence of previously unseen classes. These models are valuable in many areas, ranging from social network and food adulteration analyses to biology, where an evolving population may be present. In this paper, we focus on a two-stage Bayesian semiparametric novelty detector, also known as Brand, recently introduced in the literature. Leveraging on a model-based mixture representation, Brand allows clustering the test observations into known training terms or a single novelty term. Furthermore, the novelty term is modeled with a Dirichlet Process mixture model to flexibly capture any departure from the known patterns. Brand was originally estimated using MCMC schemes, which are prohibitively costly when applied to high-dimensional data. To scale up Brand applicability to large datasets, we propose to resort to a variational Bayes approach, providing an efficient algorithm for posterior approximation. We demonstrate a significant gain in efficiency and excellent classification performance with thorough simulation studies. Finally, to showcase its applicability, we perform a novelty detection analysis using the openly-available \texttt{Statlog} dataset, a large collection of satellite imaging spectra, to search for novel soil types.
}

\section{Introduction}

\label{sec:intro}
Statistical methods for novelty detection are becoming increasingly popular in recent literature. Similar to standard supervised classifiers, these models are trained on a fully labeled dataset and subsequently employed to group unlabeled data. In addition, novelty detectors allow for the discovery and consequent classification of samples showcasing patterns not previously observed in the learning phase.
For example, novelty detection models have been successfully employed in uncovering unknown adulterants in food authenticity studies \citep{Cappozzo2019e,Fop2022}, collective anomalies in high energy particle physics \citep{Vatanen}, and novel communities in social network analyses \citep{Bouveyron}.\\
Formally, we define a \emph{novelty detector} as a classifier trained on a set (the training set) characterized by a specific number of classes that is used to predict the labels of a second set (the test set). For a detailed account of the topic, the interested reader is referred to the reviews of \cite{Markou2003b} and \cite{Markou2003}, where the former is devoted explicitly to statistical methods for novelty detection.

Within the statistical approach, different methodologies have been proposed to construct novelty detectors. Recently, \cite{denti2021two} introduced a Bayesian Robust Adaptive Novelty Detector (Brand). Brand is a semiparametric classifier divided into two stages, focused on training and test datasets, respectively. 
In the first stage, a robust estimator is applied to each class of the labeled training set. By doing so, Brand recovers the representative traits of each known class. The second step consists in fitting a semiparametric nested mixture to the test set: the hierarchical structure of the model specifies a convex combination between terms already observed in the training set and a novelty term, where the latter is decomposed into a potentially unbounded number of novel classes. At this point, Brand uses the information extracted from the first phase to elicit reliable priors for the known components.

In their article, \cite{denti2021two} devised an MCMC algorithm for posterior estimation. The algorithm is based on a variation of the slice sampler \citep{Kalli2011} for nonparametric mixtures, which avoids the ex-ante specification of the number of previously unseen components, reflecting the expected ignorance about the structure of the novelty term. As a result, the algorithm obtains good performance as it targets the true posterior without resorting to any truncation. Nonetheless, as is often the case when full MCMC inference is performed, it becomes excessively slow when applied to large multidimensional datasets.\\
In this paper, we aim to scale up the applicability of Brand by adopting a  variational inference approach, vastly improving its computational efficiency. Variational inference is an estimating technique that approximates a complex probability distribution by resorting to optimization \citep{Jordan1999, Ormerod2010}, which has received significant attention within the statistical literature in the past decade \citep{Blei2017}. In the Bayesian set-up, the distribution of interest is, of course, the posterior distribution. Variational inference applied to Bayesian problems (also known as variational Bayes, VB from now on) induces a paradigm shift in the approximation of a complicated posterior: we switch from a simulation problem (MCMC) to an optimization one. Broadly speaking, the VB approach starts with the specification of a class of simple distributions called the variational family. Then, within this specified family, one looks for the member that minimizes a suitable distributional divergence (e.g., the Kullbak-Lielber divergence) from the actual posterior. By dealing with a minimization problem instead of a simulation one, we can considerably scale up the applicability of Brand, obtaining results for datasets with thousands of observations measured in high-dimension in a fraction of the time needed by MCMC techniques. 
Variational techniques have showcased the potential to enhance the relevance of the Bayesian framework even to large datasets, sparking interest in its theoretical properties \citep[e.g.,][]{Wang2012a,Nieman2022,Ray2022}, and its applicability to, for instance, logistic regression \citep{Jaakkola2000,Rigon}, network analysis \citep{Aliverti2022}, and, more in general, non-conjugate \citep{Wang2012} and advanced models \citep{Zhang2019}.

The present paper is structured as follows. In Section \ref{sec:background}, we review Brand and provide a summary of the variational Bayes approach. In Section \ref{sec:algo}, we discuss the algorithm and the hyperparameters needed for the VB version of the considered model. Section \ref{sec:simstud} reports extensive simulation studies that examine the efficiency and robustness of our method. Then, in Section \ref{sec:appl}, we present an application to the \texttt{Statlog} dataset, openly available from the UCI dataset repository. In this context, novelty detection is used to discover novel types of soil starting from satellite images. We remark that this analysis would have been prohibitive with a classical MCMC approach. Lastly, Section \ref{sec:disc} concludes the manuscript.

\section{Background: Bayesian novelty detection and variational Bayes}
\label{sec:background}
This section introduces the two stages in which Brand articulates and briefly presents the core concepts of the variational Bayes approach.

\subsection{The Brand model}

In what follows, we review the two-stage procedure proposed in \cite{denti2021two}. The first stage centers around a fully labeled learning set from which we extract robust information to set up a Bayesian semiparametric model in the second stage.
More specifically, consider a labeled training dataset with $n$ observations grouped into $J$ classes. This paper will focus on classes distributed as multivariate Gaussians, but one can readily extend the model using different distributions. To this regard, we will write $\mathcal{N}\left(\cdot\mid\boldsymbol{\Theta}\right)$ to indicate a Multivariate Gaussian density with generic location and scale parameters $\boldsymbol{\Theta}=\left(\boldsymbol{\mu},\Sigma\right)$. Within each of the $J$ training classes, we separately apply the Minimum Regularized Covariance Determinant (MRCD) estimator \citep{Boudt2020} to retrieve robust estimates for the associated mean vector and covariance matrix. In detail, the MRCD searches for a subset of observations with fixed size whose regularized sample covariance has the lowest possible determinant. MRCD estimates are then defined as the multivariate location and regularized scatter based on the so-identified subset. Compared to the popular Minimum Covariance Determinant \citep{Driessen1999, Hubert2018}, MRCD takes advantage of regularization with a pre-specified positive definite target matrix to ensure the existence of the solution even when the data dimension exceeds the sample size. This feature implies that the robustness properties of the original procedure are preserved whilst being applicable also to ``$p>n$'' problems. Such a characteristic is paramount in the context considered in the present paper, where we specifically aim to scale up the applicability of Brand to high-dimensional scenarios. We will denote the estimates obtained for class $j$, $j=1,\ldots,J,$ with $\bar{\Theta}^{MRCD}_j = \left(\hat {\boldsymbol{\mu}}^{MRCD}_j,\hat{\boldsymbol{\Sigma}}^{MRCD}_j\right)$. These quantities will be employed to elicit informative priors for the $J$ known components in the second stage. Specifically, in so doing, outliers and label noise that might be present in the training set will not hamper the Bayesian mixture model specification hereafter reported.

In the second stage, we estimate a semiparametric Bayesian classifier on a test set with $M$ unlabeled observations. We want to build a novelty detector, i.e., a model that can discern between ``known'' units - which follow a pattern already observed in the training set - or ``novelties''. At this point, the likelihood for each observation is a simple two-group mixture between a generic novelty component $f_{nov}$ and a density reflecting the behavior found in the training set $f_{obs}$. Given stage 1, it is immediate to specify $f_{obs}$ as a mixture of $J$ multivariate Gaussians, whose hyperpriors elicitation will be guided by each of the robust estimates $\bar{\Theta}^{MRCD}_j$. Therefore, we can now write the likelihood for the generic observation $\boldsymbol{y}_m$, where $m=1,\ldots, M$, as a mixture of $J+1$ distributions:
\begin{equation}
    \mathcal{L}\left(\boldsymbol{y}_m \mid \boldsymbol{\Theta}^{obs}, \boldsymbol{\pi}\right) = \pi_0 f_{nov} + \sum_{j=1}^{J} \pi_j\mathcal{N}\left(\boldsymbol{y}_m\mid \boldsymbol{\Theta}^{obs}_{j}\right).
    \label{eq:genlik}
\end{equation}
Estimating this model allows us to either allocate each of the $M$ observations into one of the previously observed $J$ Gaussian classes or flag it as a novelty generated from an unknown distribution $f_{nov}$.

The specification of $f_{nov}$ is more delicate. To specify a flexible distribution that would reflect our ignorance regarding the novelty term, we employ a Dirichlet process mixture model with Gaussian kernels \citep{Escobar1995}. It is a mixture model where the mixing distribution is sampled from a Dirichlet process \citep{Ferguson1973}, characterized by concentration parameter $\gamma$ and base measure $H$. In other words, we have that $f_{nov}(\boldsymbol{y}_m) = \sum_{k=1}^\infty \omega_k \mathcal{N}\left(\boldsymbol{y}_m \mid \boldsymbol{\Theta}^{nov}_{k}\right)$, where the atoms are sampled from the base measure, i.e., $\boldsymbol{\Theta}^{nov}\sim H$, and the mixture weights follow the so-called stick-breaking construction \citep{Sethuraman1994a}, where $w_k = v_k\prod_{l<k}(1-v_l)$ and $v_l\sim Beta(1,\gamma)$. To indicate this process, we will write $\boldsymbol{\omega}\sim SB(\gamma)$. Thus, the likelihood in \eqref{eq:genlik} becomes, for $m=1,\ldots,M$,
\begin{equation}
     \mathcal{L}\left(\boldsymbol{y}_m \mid \boldsymbol{\Theta}, \boldsymbol{\pi}\right) = \pi_0 \left[\sum_{k=1}^\infty \omega_k \mathcal{N}\left(\boldsymbol{y}_m \mid \boldsymbol{\Theta}^{nov}_{k}\right)\right] + \sum_{j=1}^{J} \pi_j\mathcal{N}\left(\boldsymbol{y}_m \mid\boldsymbol{\Theta}_{j}\right).
     \nonumber
\end{equation}
This nested-mixture expression of the likelihood highlights a two-fold advantage. First, it is highly flexible, effectively capturing departures from the known patterns and flagging them as novelties. Second, the mixture nature of $f_{nov}$ allows to automatically cluster the novelties, capturing potential patterns that may arise. Furthermore, clusters in the novelty terms characterized by very small sizes can be interpreted as simple outliers.

Notice that the previous nested-mixture likelihood can be conveniently re-expressed as:
\begin{align}
    \mathcal{L}\left(\boldsymbol{y}_m \mid \boldsymbol{\Theta}^{obs}, \boldsymbol{\pi}\right) &= \sum_{k=1}^{\infty} \tilde{\pi}_k \mathcal{N}\left(\boldsymbol{y}_m\mid \tilde{\boldsymbol{\Theta}}_{k}\right),
    \label{eq:marg_lik}
\end{align}
where, for $k \in \mathbb{N}$, we define $\tilde\pi_k = {\pi_k}^{\mathbbm{1}{\{0 < k \leq J\}}}(\pi_0\omega_{k-J})^{\mathbbm{1}{\{k \geq J\}}}$ and 
$\tilde{\boldsymbol{\Theta}}_k = (\boldsymbol{\Theta}_k^{obs})^{\mathbbm{1}{\{0 < k \leq J\}}}(\boldsymbol{\Theta}_{k}^{nov})^{\mathbbm{1}{\{k \geq J\}}}$. Note that, without loss of generality, we regard the first $J$ mixture components as the known components and all the remaining ones as novelties.

Finally, as customary in mixture models, to ease the computation, we augment the model by introducing the auxiliary variables $\xi_m\in \mathbb{N}$, for $m=1,
\ldots,M$, where $\xi_m = l$ means that the $m$-th observation has been assigned to the $l$-th component. Therefore, the model in \eqref{eq:marg_lik} simplifies to
\begin{align}
    \boldsymbol{y}_m \mid \xi_m,\tilde{\boldsymbol{\Theta}}, \boldsymbol{\pi}\sim & \mathcal{N}\left(\boldsymbol{y}_m\mid \tilde{\boldsymbol{\Theta}}_{\xi_m}\right),\quad \quad
    {\xi_m}\mid{\tilde{\boldsymbol{\pi}} \overset{iid}{\sim}  \sum_{k=1}^\infty{\tilde\pi_k\delta_k(\cdot)} }.
    \label{eq:aux_spec}
\end{align}
We complete our Bayesian model with the following prior specifications for the weights and the atoms:
\begin{equation}
    \begin{aligned}
    \boldsymbol{\pi} &\sim {Dirichlet}(\alpha_0, \alpha_1, ..., \alpha_J),\quad \quad 
     \boldsymbol{\omega} \sim {SB}(\gamma),\\
    \boldsymbol{\Theta}_k^{obs} &\sim {\mathcal{NIW}}(\boldsymbol{\mu}^{{obs}}_k, \nu^{{obs}}_k, \lambda^{{obs}}_k, \boldsymbol{\Psi}^{{obs}}_k), \quad \quad k = 1,\ldots,J,\\
    \boldsymbol{\Theta}_k^{nov} &\sim  {\mathcal{NIW}}({\boldsymbol{\mu}_0^ {nov}}, \nu_0^{ {nov}}, \lambda_0^{ {nov}}, \boldsymbol{\Psi}_0^{nov}),  \quad \quad k = J+1,\ldots, \infty,
    \label{eq:prior_NIW}
\end{aligned}
\end{equation}
where $\mathcal{NIW}$ indicates a Normal Inverse-Wishart distribution. To ease the notation, let $\boldsymbol{\Theta} = \{\boldsymbol{\Theta}_k^{obs}\}_{k=1}^J \cup \{\boldsymbol{\Theta}_k^{nov}\}_{k=J+1}^\infty$, $\boldsymbol{\varrho}_k = (\boldsymbol{\mu}^{{obs}}_k, \nu^{{obs}}_k, \lambda^{{obs}}_k, \boldsymbol{\Psi}^{{obs}}_k)$ and $\boldsymbol{\varrho}_0 = (\boldsymbol{\mu}^{nov}_0, \nu^{nov}_0, \lambda^{nov}_0, \boldsymbol{\Psi}^{nov}_0)$. We remark that the values for the hyperparameters $\{\boldsymbol{\varrho}_k^{obs}\}_{k = 1}^J$ are defined according to the robust estimates obtained in the first stage. See \cite{denti2021two} for more details about the hyperparameters specification.

\begin{figure}[t!]
    \centering
    \includegraphics[width = \linewidth]{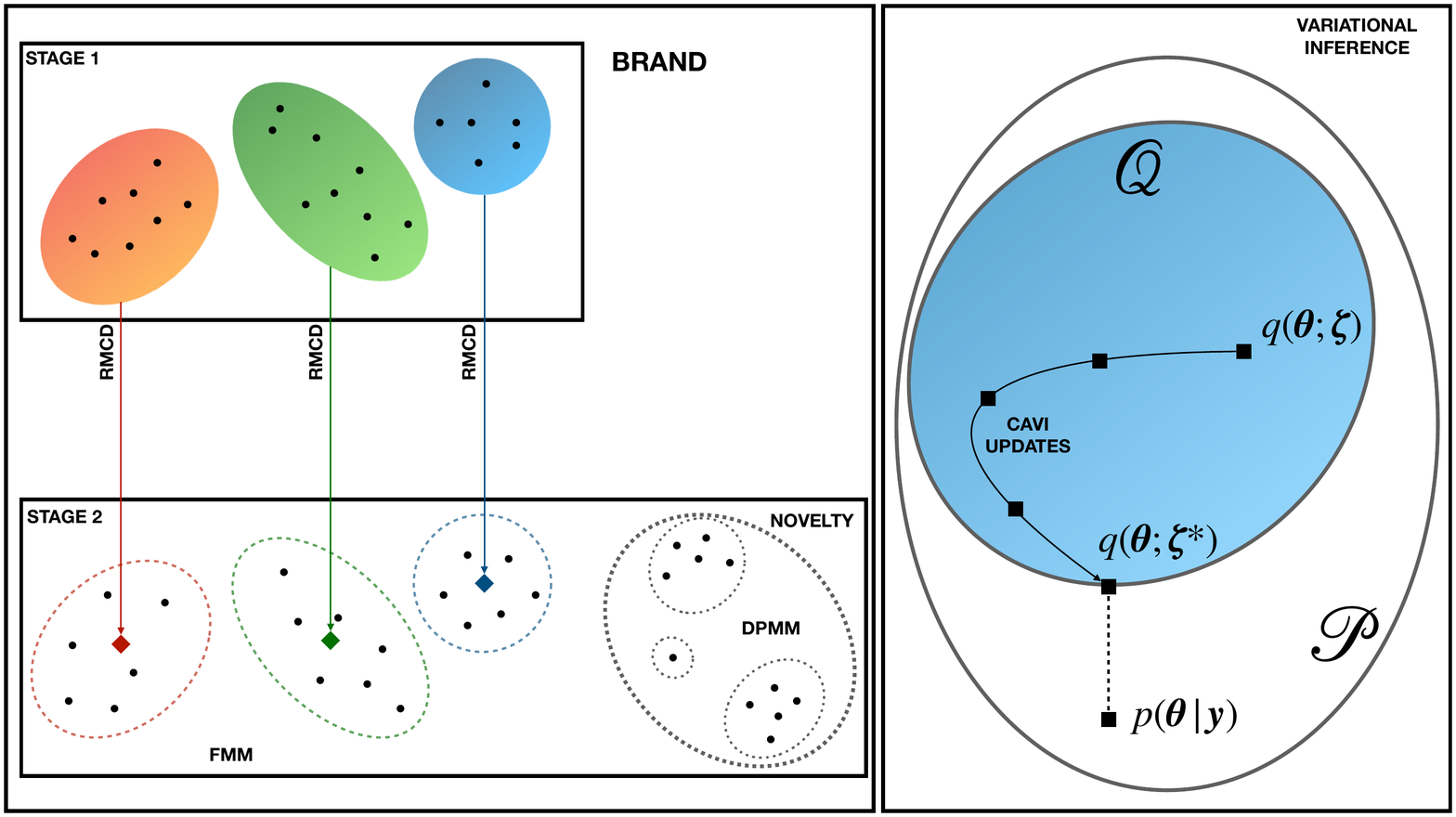}
    \caption{Diagrams depicting how Brand (left panel) and variational inference (right panel) work, respectively.}
    \label{fig:brand_vb}
\end{figure}

Given the model, it is easy to derive the joint law of the data and the parameters, which is proportional to the posterior distribution of interest $p\left({\boldsymbol{\Theta}}, \boldsymbol{\pi},\boldsymbol{\xi} \mid \boldsymbol{y} \right)$.
Therefore, the posterior we target for inference is proportional to:
\begin{align}
 p\left({\boldsymbol{\Theta}}, \boldsymbol{\pi},\boldsymbol{\xi} \mid \boldsymbol{y} \right)  \propto & 
 \prod_{m=1}^M\left[ \prod_{k=1}^J\mathcal{N}(\boldsymbol{y}_m\mid\boldsymbol\Theta_k^{obs})^{\mathbbm{1}_{\xi_m=k}}\prod_{k=J+1}^{\infty}\mathcal{N}(\boldsymbol{y}_m\mid\boldsymbol\Theta_k^{nov})^{\mathbbm{1}_{\xi_m=k}}\right] \times 
 \nonumber \\ & \prod_{k=1}^J\mathcal{\mathcal{NIW}}(\boldsymbol\Theta_k^{obs}\mid \boldsymbol{\varrho}_k)  \prod_{k=J+1}^{\infty}{\mathcal{NIW}}(\boldsymbol\Theta_k^{nov}\mid\boldsymbol{\varrho}_0)\times \nonumber \\ & \prod_{m=1}^M \left[ \prod_{k=1}^J(\pi_k)^{\mathbbm{1}_{\xi_m=k}}\prod_{k=J+1}^{\infty}\pi_0\left( v_{k-J}\prod_{h=1}^{k-J-1}(1-v_h)\right)^{\mathbbm{1}_{\xi_m=k}}\right] \times \nonumber 
  \\
  & \prod_{k=0}^J(\pi_k)^{\alpha_k-1}\prod_{k=1}^{\infty}(1-v_k)^{\gamma-1}.
  \label{eq:post}
 \end{align}
The left panel of Figure \ref{fig:brand_vb} contains a diagram that summarizes how Brand works.
In the following, we will devise a VB approach to approximate \eqref{eq:post} in a timely and efficient manner. The following subsection briefly outlines the general strategy underlying a mean-field variational approach, while the thorough derivation of the algorithm used to estimate Brand is deferred to Section \ref{sec:algo}.

\subsection{A short summary of mean-field variational Bayes}

Working in a Bayesian setting, we ultimately seek to estimate the posterior distribution to draw inference. Unfortunately, the expression in \eqref{eq:post} is not available in closed form, and therefore we need to rely on approximations. 
MCMC algorithms, which simulate draws from \eqref{eq:post},  could be prohibitively slow when applied to large datasets. To this extent, we resort to variational inference to recasts the approximation of \eqref{eq:post} into an optimization problem. For notational simplicity, while we present the basic ideas of VB, let us denote a generic target distribution with $p(\boldsymbol{\theta}\mid \boldsymbol{y})\propto p(\boldsymbol{\theta}, \boldsymbol{y})$. 

As in \cite{Blei2017}, we focus on a mean-field variational family $\mathcal{Q}$, a set containing distributions for the parameters of interest $\boldsymbol{\theta}$ that are all mutually independent: $\mathcal{Q} = \{q_{\boldsymbol{\zeta}}(\boldsymbol{\theta}): q_{\boldsymbol{\zeta}}\boldsymbol{\theta} = \prod_jq_{\zeta_j} ({\theta}_j)\}$. The postulated independence dramatically simplifies the problem at hand. Notice that each member of $\mathcal{Q}$ depends on a set of variational parameters denoted by $\boldsymbol{\zeta}$.

We seek, among the members of this family, the candidate that provides the best approximation of our posterior distribution $p(\boldsymbol{\theta}\mid \boldsymbol{y})$. Herein, the goodness of the approximation is quantified by the Kullback-Leibler (KL) divergence. Thus, we aim to find the member of the variational family $\mathcal{Q}$ that minimizes the KL divergence between the actual posterior distribution and the variational approximation. 
The KL divergence ${D}_{KL} (\cdot\mid\mid\cdot)$ can be written as ${D}_{KL} (p(\boldsymbol{\theta} \mid \boldsymbol{y})\mid\mid q_{\boldsymbol{\zeta}}(\boldsymbol{\theta}))) = \mathbb{E}_q[\log{q_{\boldsymbol{\zeta}}(\boldsymbol{\theta})}] - \mathbb{E}_q[\log{p(\boldsymbol{\theta},\boldsymbol{y})}]+\log p(\boldsymbol{y}).$ Unfortunately, we cannot easily compute the \emph{evidence} $p(\boldsymbol{y})$. However, the evidence does not depend on any variational parameter and can be treated as fixed w.r.t. $\boldsymbol{\theta}$ during the optimization process. We then re-formulate the problem into an equivalent one, the maximization of the \textit{Evidence Lower Bound} (ELBO), which is fully computable:
\begin{equation}
 ELBO(q)  = \mathbb{E}_q[\log{p(\boldsymbol{z},\boldsymbol{y})}] - \mathbb{E}_q[\log{q_{\boldsymbol{\zeta}}(\boldsymbol{z})}].
 \label{eq:elbo}
\end{equation}
Maximizing \eqref{eq:elbo} is equivalent to minimize the aforementioned KL divergence.\\
To detect the optimal member $q^\star\in \mathcal{Q}$, we employ a widely used algorithm called \emph{Coordinate Ascent Variational Inference} (CAVI). It consists of a one-variable-at-a-time optimization procedure. Indeed, exploiting the independence postulated by the mean-field property, one can show that the optimal variational distribution for the parameter $\theta_j$ is given by:
\begin{equation}
    q^\star_{\zeta_j}(\theta_j)\propto
    \exp\{\mathbb{E}_{-j}\left[\log p(\theta_j, \boldsymbol{\theta}_{-j},\boldsymbol{y})\right]\},
    \label{eq:step}
\end{equation}
where $\boldsymbol{\theta}_{-j} = \{\theta_l\}_{l\neq j}$ and the expected value is taken w.r.t. the densities of $\boldsymbol{\theta}_{-j}$. The CAVI algorithm iteratively computes \eqref{eq:step} for every $j$ until the ELBO does not register any significant improvement. The basic idea behind the CAVI algorithm is depicted in the right-half of Figure \ref{fig:brand_vb}.\\ 
In the next subsections, we will derive the CAVI updates for the variational approximation of our model, along with the corresponding expression of the ELBO.
We call the resulting algorithm the variational Brand, or \texttt{VBrand}.

\section{Variational Bayes for the Brand model}
\label{sec:algo}

In this section, we tailor the generic variational inference algorithm to our specific case. First, we write our variational approximation, highlighting the dependence of each distribution on its specific variational parameters, collected in $\boldsymbol{\zeta} =\left( \boldsymbol{\eta},\boldsymbol{\varphi},\boldsymbol{a},\boldsymbol{b},\boldsymbol{\rho}^{obs},\boldsymbol{\rho}^{nov}
\right)$. The factorized form we adopt is:
\begin{equation}\begin{split}
 q_{\boldsymbol{\zeta}}( \boldsymbol{\xi}, \boldsymbol{\pi}, \boldsymbol{v}, \boldsymbol{\Theta}^{obs}, \boldsymbol{\Theta}^{nov})  =& q_{\boldsymbol{\eta}}(\boldsymbol{\pi})  \prod_{m=1}^M q_{\boldsymbol{\varphi}^{(m)}}(\xi^{(m)})  \prod_{k=1}^{T-1} q_{a_k,b_k}(v_k)\times \\  
  &\prod_{k=1}^J q_{\boldsymbol{\rho}_k^{obs}} (\boldsymbol{\Theta}^{obs}_k) 
 \prod_{k=J+1}^{J+T} q_{\boldsymbol{\rho}_k^{nov}} (\boldsymbol{\Theta}^{nov}_k).
 \end{split}
 \label{eq:Vfam}
 \end{equation}
In Equation \eqref{eq:Vfam}, we truncated the stick-breaking representation of the Dirichlet Process on the novelty term at a pre-specified threshold $T$, as suggested in \cite{Blei2006}. This implies that $q(v_T = 1) = 1$ and that all the variational mixture weights indexed by $t > T $ are equal to zero.

Then, we can exploit a key property of VB. Note that all the full conditionals of the Brand model have closed-form expressions (c.f.r. Section S1 of the Supplementary Material) and belong to the exponential family. This feature greatly simplifies the search for the variational solution. Indeed, it ensures that the corresponding optimal variational distributions belong to the same family of the corresponding full conditional, with properly updated variational parameters. Therefore, we can already state that $q_{\boldsymbol{\eta}}(\boldsymbol{\pi})$ is the density function of a Dirichlet distribution, $q_{\boldsymbol{\varphi}^{(m)}}(\xi^{(m)})$ follows a Categorical distribution, 
each $q_{a_k,b_k}(v_k)$ follows a Beta distributions, and
$q_{\boldsymbol{\rho}_k^{obs}} (\boldsymbol{\Theta}^{obs}_k)$ and $q_{\boldsymbol{\rho}_k^{nov}} (\boldsymbol{\Theta}^{nov}_k)$ are both distributed according to a Normal Inverse-Wishart.\\

\subsection{The CAVI parameters update}

Once the parametric expressions for the members of the variational family are obtained, we can derive the explicit formulas to optimize the parameters via the CAVI algorithm. In this subsection, we state the updating rules that have to be iteratively computed to fit \texttt{VBrand}. 
As we will observe, the set of responsibilities $\{\varphi_k^{(m)}\}_{k = 1}^{J+T}$, i.e., the variational probabilities $\varphi^{(m)}_k=q(\xi^{(m)}=k)$, will play a central role in all the steps. In detail, the CAVI steps are as follows:
\begin{enumerate}
    \item $q^\star_{\boldsymbol{\eta}}(\boldsymbol{\pi})$ is the density of a $Dirichlet(\eta_0,\eta_1,\ldots,\eta_J)$, where
\begin{equation}
\eta_0 = 
    \alpha_0 + \sum_{m=1}^M\sum_{l=J+1}^{J+T}\varphi_l^{(m)},
        \quad \;
     \eta_j = \alpha_j + \sum_{m=1}^M\varphi_j^{(m)},  \; \text{ for }j=1,\ldots,J.\\
     \label{eq:upd_eta}
\end{equation}

Here, each hyperparameter linked to a known component is updated with the sum of the responsibilities of the data belonging to the same specific known component. Likewise, the variational novelty probability hyperparameter $\eta_0$ contains the sum of the responsibilities of all data belonging to all the novelty terms.\\

\item  Each $q^\star_{a_k,b_k}(v_k)$, for $k=1,\ldots,T-1$, is the density of a $Beta(a_k,b_k)$. The update for these variational parameters is given by:
\begin{equation}
a_k = 1 + \sum_{m=1}^M\varphi_{J+k}^{(m)}, \quad b_k = \gamma + \sum_{l=k+1}^T\sum_{m=1}^M\varphi_{J+l}^{(m)}.
\label{eq:upd_ab_beta}
\end{equation}
Here, as expected from the stick-breaking construction, the first parameter of the variational $Beta$ distribution is updated with the sum of the probabilities of each point belonging to a specific novelty cluster $k$. At the same time, the second parameter is updated with the sum of variational probabilities of belonging to one of the next novelty components.\\

\item Both $q^\star_{\boldsymbol{\rho}_k^{nov}} (\boldsymbol{\Theta}^{nov}_k)$ and
$q^\star_{\boldsymbol{\rho}_k^{obs}} (\boldsymbol{\Theta}^{obs}_k)$ are Nomal Inverse-Wishart densities. Let us start with the updating rules of the known components.
Note that each variational parameter $\boldsymbol{\rho}_k^{obs}$ is a shorthand for $\left(\boldsymbol{m}^{obs}_k,\ell^{obs}_k, u^{obs}_k, \boldsymbol{S}^{obs}_k \right)$. These parameters have the same interpretation as the parameters of \eqref{eq:prior_NIW}, contained in $\boldsymbol{\varrho}_k$. So we have
\begin{equation}
\begin{aligned}
\boldsymbol{m}^{obs}_k &= \frac{1}{\lambda^{obs}_k+\sum_{m=1}^M\varphi_{k}^{(m)}}\left(\lambda^{obs}_k\boldsymbol{\mu}^{obs}_k + \sum_{m=1}^M\boldsymbol{y}_m\varphi_{k}^{(m)}\right), \\
\ell_k &= \lambda_k^{obs} + \sum_{m=1}^M\varphi_{k}^{(m)},\quad
u^{obs}_k = \nu_K^{obs} + \sum_{m=1}^M\varphi_{k}^{(m)},\\
\boldsymbol{S}^{obs}_k &= \boldsymbol{\Psi}^{obs}_k + \sum_{m=1}^M \hat{\boldsymbol{\Sigma}}_{k}^{(m)} + \frac{\lambda^{obs}_k\sum_{m=1}^M\varphi_{k}^{(m)}}{\lambda^{obs}_k+\sum_{m=1}^M\varphi_{k}^{(m)}}(\overline{\boldsymbol{y}}_k-\boldsymbol{\mu}^{obs}_k)^T(\overline{\boldsymbol{y}}_k-\boldsymbol{\mu}^{obs}_k),
\label{eq:upd_NIW}
\end{aligned}
\end{equation}
where we defined $\hat{\boldsymbol{\Sigma}}_{k}^{(m)} = (\boldsymbol{y}_m-\overline{\boldsymbol{y}}_k)(\boldsymbol{y}_m-\overline{\boldsymbol{y}}_k)^T\varphi_{k}^{(m)}$ and
$\overline{\boldsymbol{y}}_k = \sum_{m=1}^M\boldsymbol{y}_m\varphi_{k}^{(m)}/\sum_{m=1}^M\varphi_{k}^{(m)}.$ The update for the parameters in $\boldsymbol{\rho}_k^{nov}$ follows the same structure, with the hyperprior parameters in $\boldsymbol{\varrho}^{nov}$ carefully substituted to $\boldsymbol{\varrho}^{obs}$.\\ 

\item Updating the responsibilities $\{\varphi_k^{(m)}\}_{k = 1}^{J+T}$ for $m=1,\ldots,M,$ is the most challenging step of the algorithm, given the nested nature of the mixture in \eqref{eq:marg_lik}. We recall that, for a given $m$, the distribution $q_{\boldsymbol{\varphi}^{(m)}}(\xi^{(m)})$ is categorical with $J+T$ levels. Thus, we need to compute the values for the $J+T$ corresponding probabilities. For the known classes $k=1,\ldots,J$, we have
\begin{equation}
 \log \varphi_k^{(m)} \propto 
       \mathbb{E}[\log{\pi_k}] + \mathbb{E}[\log{\mathcal{N}(\boldsymbol{y}_m\mid\boldsymbol{\Theta}_k^{obs})}],  
             \label{eq:upd_obs}
  \end{equation}
while for the novelty terms $k=J+1,\ldots,J+T$, we have 
\begin{equation}
\begin{aligned}
  \log\varphi_k^{(m)} \propto \:\: &   \mathbb{E}[\log{\pi_0}] + \mathbb{E}[\log{v_{k-J}}] + \sum_{l=1}^{k-J-1}\mathbb{E}[\log\left(1-v_{l}\right)] + \\ & \mathbb{E}[\log{\mathcal{N}(\boldsymbol{y}_m\mid\boldsymbol{\Theta}_{k}^{nov})}].
      \label{eq:upd_nov}
      \end{aligned}
\end{equation}
For the sake of conciseness, we report the explicit expression for all the terms of \eqref{eq:upd_obs} and \eqref{eq:upd_nov} in Section S2 of the Supplementary Material.

This means that the probability of datum $\boldsymbol{y}_m$ to belong to cluster $k$ depends on the likelihood of $\boldsymbol{y}_m$ under that same cluster and on the overall relevance of $k^{th}$ cluster. Such relevance is determined as the expected value of the relative component of the ${Dirichlet}$ distributed $\boldsymbol\pi$ and, for novelties, the stick-breaking weight, here unrolled in its ${Beta}$-distributed components.\\

\end{enumerate}

\subsection{The expression of the ELBO for \texttt{VBrand}}

In this subsection, we report the terms that need to be derived to obtain the ELBO in Equation \eqref{eq:elbo} for \texttt{VBrand} model. We start by computing the first term of Equation \eqref{eq:elbo}, which takes the following form:
\begin{equation*}\begin{split}
 \mathbb{E}[\log{p}] = & \sum_{m=1}^M\left(\sum_{k=1}^J f_{1}^{(m,k)} + \sum_{k=J+1}^{J+T} f_{2}^{(k)} \right) + \sum_{k=1}^J f_{3}^{(k)}  + \sum_{k=J+1}^{J+T} f_{4}^{(k)} +\\&  \sum_{m=1}^M\left(\sum_{k=1}^J f_{5}^{(m,k)} + \sum_{k=J+1}^{J+T} f_{6}^{(m,k)} \right) + \sum_{k=0}^{J} f_{7}^{(k)} + \sum_{l=1}^{T} f_{8}^{(l)} + const,
 \end{split}\end{equation*}
where the quantities $\{f_k\}_{k=1}^8$ have the following expressions (note we have suppressed the superscripts to ease the notation, and that $\boldsymbol{\psi}(\cdot)$ indicates the digamma function):
\begin{equation*}
    \begin{aligned}
 f_{1} &= \varphi_{k}^{(m)}\mathbb{E}[\log{{N}(\boldsymbol{y}_m\mid\boldsymbol{\Theta}_k^{obs})}],&\quad
 f_{2} &= \varphi_{k}^{(m)}\mathbb{E}[\log{{N}(\boldsymbol{y}_m\mid\boldsymbol{\Theta}_k^{nov})}], \\
 f_{3} &= \mathbb{E}[\log{{\mathcal{NIW}}(\boldsymbol{\Theta}_k^{obs}\mid \boldsymbol{\varrho}_k)}], & \quad f_{4} &= \mathbb{E}[\log{{\mathcal{NIW}}(\boldsymbol{\Theta}_k^{nov}\mid \boldsymbol{\varrho}_0})].
    \end{aligned}
\end{equation*}
Moreover, we have $f_{5} = \varphi_{k}^{(m)}\left(\boldsymbol{\psi}\left(\eta_k\right)-\boldsymbol{\psi}\left(\sum_{j=0}^J\eta_j\right)\right)$,
\begin{equation*}
\begin{aligned}
 f_{6} = \varphi_{k}^{(m)}\left[\boldsymbol{\psi}(\eta_0)-\boldsymbol{\psi}(\sum_{j=0}^J\eta_j) +  \boldsymbol{\psi}(a_{k-J}+b_{k-J}) + \sum_{h=1}^{k-J-1}\boldsymbol{\psi}(b_{h})-\boldsymbol{\psi}(a_{h}+b_{h})\right],
 \end{aligned}
\end{equation*}
and, lastly, $f_{7} = (\alpha_k-1)(\boldsymbol{\psi}(\eta_k)-\boldsymbol{\psi}(\sum_{j=0}^J\eta_j)),$ and $f_{8} = (\gamma-1)(\boldsymbol{\psi}(b_l)-\boldsymbol{\psi}(a_l+b_l))$. \\
The second term of Equation \eqref{eq:elbo} can be written as
\begin{equation*}
  \mathbb{E}[\log{q}] = \sum_{m=1}^{M}\left(h_1^{(m)} + h_2^{(m)} + \sum_{k=1}^{T}h_3^{(m,k)} + \sum_{k=1}^{J}h_4^{(m,k)} + \sum_{k=1}^{T}h_5^{(m,k)}\right),
\end{equation*}
with $h_{1} = \sum_{k=1}^{J+T}\varphi_{k}^{(m)}\ln{\varphi_{k}^{(m)}} - \ln{\sum_{k=0}^{J+T}\varphi_{k}^{(m)}}$, 
$h_{4} = \mathbb{E}[\log{\mathcal{NIW}(\boldsymbol{\Theta}_k^{obs}\mid \boldsymbol{\rho}_k^{obs})}]$, \\
$h_{5}^{(m,k)} = \mathbb{E}[\log{\mathcal{NIW}(\boldsymbol{\Theta}_k^{nov}\mid \boldsymbol{\rho}_k^{nov})}]$, and, finally,
\begin{equation*}
\begin{aligned}
 h_{2} =& \sum_{j=0}^{J}(\eta_j-1)\left(\boldsymbol{\psi}(\eta_j)-\boldsymbol{\psi}(\sum_{j=0}^J\eta_j)\right) +
 \ln{\Gamma\left(\sum_{j=0}^J{\eta_j}\right)} - \sum_{j=0}^{J}\ln{\Gamma(\eta_j)},\\
 h_{3} =& (a_k-1)(\boldsymbol{\psi}(a_k)-\boldsymbol{\psi}(a_k+b_k)) +\\ & (b_k-1)(\boldsymbol{\psi}(b_k)-\boldsymbol{\psi}(a_k+b_k)) - \ln{\left(\frac{\Gamma(a_k)\Gamma(b_k)}{\Gamma(a_k+b_k)}\right)}.
 \end{aligned}
\end{equation*}
Additional details are deferred to Section S2 of the Supplementary Material. \\

At each step of the algorithm, one needs to update the parameters according to the rules \eqref{eq:upd_eta}-\eqref{eq:upd_nov} and to evaluate all the terms in $\{f_k\}_{k=1}^8$ and $\{h_k\}_{k=1}^5$. Albeit faster than MCMC, this articulates over numerous steps. To further reduce the overall computing time, an \texttt{R} package relying on an efficient \texttt{C++} implementation has been implemented. The package is openly available at the GitHub repository \url{JacopoGhirri/VarBRAND}. In the same repository, the interested reader can find all the \texttt{R} scripts written to run the simulation studies and real data application that we will discuss in the following sections. 


\section{Simulation studies}
\label{sec:simstud}
In this section, we report the performance of our variational algorithm on a range of simulated datasets. In particular, our simulation study is articulated into three different experiments. Each experiment investigates a different aspect of the model while altogether providing a multi-faceted description of its performance. 

The first experiment focuses on the scaling capabilities of our proposal; the second compares the results and efficiency of \texttt{VBrand} with the original slice sampler, while the last one assesses the sensitivity of the recovered partition to the hyperprior specification. For the MCMC algorithm, we follow the default hyperprior setting suggested in \cite{denti2021two}. Moreover, we run the slice sampler for 20,000 iterations, discarding the first 10,000 as burn-in. As for \texttt{VBrand}, we use the same hyperprior specifications used for the MCMC (unless otherwise stated), while we use $k$-means estimation to initialize the novelty terms' means. We set a threshold $\varepsilon = 10^{-9}$ as stopping rule.

\begin{table}[t!]
    \centering
    \begin{tabular}{cccccccc}
        \toprule
        SS1  & $C_1$ & $C_2$ & $C_3$ & $C_4^*$ & $C_5^*$ & $C_6^*$ & $C_7^*$ \\
        \midrule
        $\boldsymbol{\mu}_k$  & (-5,-5) & (-4,-4) & (4,4) & (0,0) & (5,-10) & (5,-10) & (-10,-10)\\
        $\sigma^2$          & 1 & 2 & 2 & 1 & 1 & 1 & 0.1\\
        $\rho$   & 0.9  & 0 & 0 & -0.75 & 0.9 & 0.9 & 0\\
        $n_{Tr}$ &  300 & 300 & 300 & -- &--& --&--\\ 
        $n_{Te}$ &  200 & 200& 250& 90& 100& 100& 60\\
        \midrule 
        SS2  & $C_1$ & $C_2$ & $C_3^*$ & $C_4^*$ & $C_5^*$ & --& -- \\
        \midrule
        $\boldsymbol{\mu}_k$  & (2,2)  & (-2,-2) & (2,-2) & (-2,2) & (0,0) & -- & --\\
        $n_{Tr}$ &  500 & 500 & -- & -- &-- & -- & --\\ 
        $n_{Te}$ &  200 & 200& 200 & 200 & 200 & -- & --\\
        \midrule
        SS3  & $C_1$ & $C_2$ & $C_3$ & $C_4^*$ & $C_5^*$ & $C_6^*$ & -- \\
        \midrule
        $\boldsymbol{\mu}_k$  & (-5,5)  & (5,5) & (5,-5) & (0,0) & (0,0) & (-5,5) & --\\
        $\sigma^2$ $^\dagger$ & 0.5 & 0.5 & 0.5 & 0.5 & 0.5 & 1.5 & -- \\
        $\rho$   & 0  & 0 & 0 & 0.8 & -0.8 & 0 & --\\
        $n_{Tr}$ &  50 & 50 & 50 & -- &-- & -- & --\\ 
        $n_{Te}$ &  1950 & 1950 & 1950 & 2000 & 2000 & 100 & --\\
        \bottomrule
    \end{tabular}
    \caption{Characteristics of the synthetic datasets used in the three simulation studies (SS1$-$SS3). The components flagged with $^*$ are novelties. $^\dagger$ the variances reported in this table refer to the low overlap scenario. For the high overlap scenario, we consider $\sigma^{2\star} = 6\sigma^2$.}
    \label{tab:simuall}
\end{table}

\subsection{Classification performance}
\label{simu::scaling}

We test the classification performance of our proposal by applying \texttt{VBrand} to a sequence of increasingly complex variations of a synthetic dataset. We monitor the computation time and different clustering metrics to provide a complete picture of the overall performance. In particular, we compute the Adjusted Rand Index (ARI, \cite{Hubert1985}), the Adjusted Mutual Information (AMI, \cite{Vinh2009}), and the Fowlkes-Mallows Index (FMI, \cite{Fowlkes1983}). While the first two metrics correct the effect of agreement solely due to chance, the latter performs well also if noise is added to an existing partition. Thereupon, the joint inspection of these three quantities aim to provide a complete picture of the results.

The considered data generating process (DGP) for this experiment is based on a mixture of 7 bivariate Normals. In detail, the first three components represent the known classes, appearing in both training and test sets, while the remaining $4$ are deemed to be novelties, present only in the test set. The components are characterized by different mean vectors $\boldsymbol{\mu}_k$, correlation matrices $(\sigma_k^2,\rho_k)$, and cardinalities in the training ($n_{Tr}$) and test ($n_{Te}$). The main attributes of this DGP are summarised in the first block, named SS1, of Table \ref{tab:simuall}. 

Starting from this basic mechanism, we subsequently increment the difficulty of the classification tasks. Specifically,  
we increase both the data dimension and the sample size as follows: 
\begin{itemize}
    \item \emph{Sample size}: while keeping unaltered the mixture proportions, we consider the sample sizes $\tilde{n}_k=q\cdot n_k$, with multiplicative factor $q\in \{0.5, 1, 2.5, 5, 10\}$ in both the training and test sets;
    \item \emph{Data dimensionality}: we augment the dimensionality $p$ of the problem by considering  $p\in \{2, 3, 5, 7, 10\}$. Each added dimension (above the second) comprises independent realization from a standard Gaussian. Note that the resulting datasets define a particularly challenging discrimination task: all the information needed to distinguish the different components is contained in the first two dimensions. In contrast, the remaining ones only display overlapping noise.
\end{itemize}
For each combination of sample size and dataset dimension, we generate 50 replications of each simulated dataset and summarize the results by computing the means and the standard errors of the chosen metrics. 
\begin{figure}[t]
    \centering
    \includegraphics[width = \linewidth]{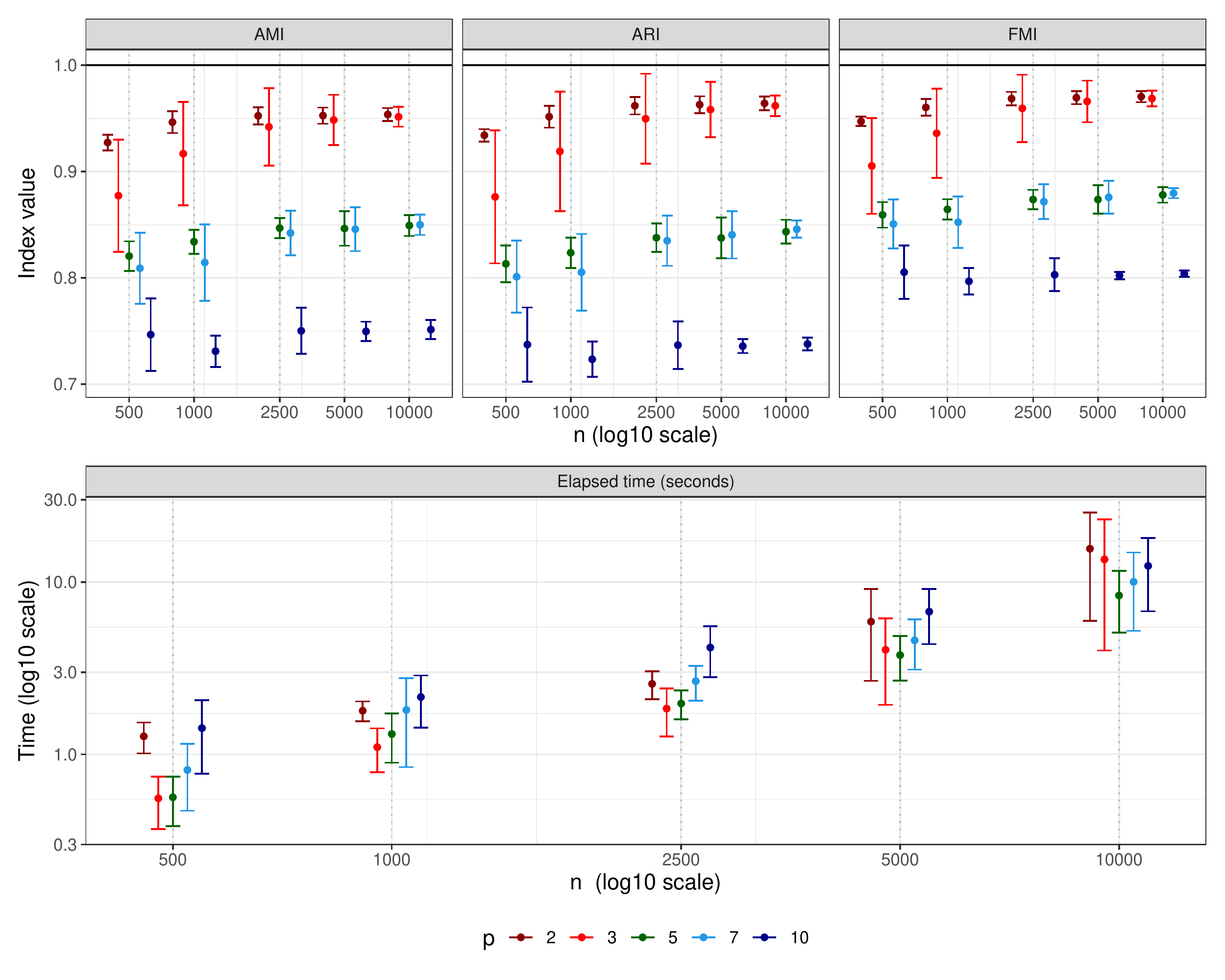}
    \caption{Performance metrics and elapsed time obtained by the \texttt{VBrand} algorithm stratified by number of variables $p$ and size scaling $n$. The dots represent the averages obtained over 50 replicates of the simulated experiment, while the vertical bars display the associated standard errors.}
    \label{fig:scaling_VB}
\end{figure}
Results for this experiment are reported in Figure \ref{fig:scaling_VB}. 
We immediately notice that the clustering performances deteriorate as the dimensionality of the problem increases. This trend is expected, especially given the induced overlap in the added dimensions. However, the classification abilities of our method remain consistent and satisfactory across all metrics. Indeed, as we can see from the three panels at the top, ARI, AMI, and FMI are all strictly above 70\% across all scenarios. This outcome indicates that not only are the known classes correctly identified and clustered as such, but the flexible nonparametric component effectively captures the novelty term.
The computation time (bottom panel) grows exponentially as a function of the test set cardinality. Interestingly, the increment of data dimensionality does not significantly impact the computational costs, suggesting an effective scalability of our proposal to high dimensional problems. Indeed, even when the test size is in the order of tens of thousands, the devised CAVI algorithm always reaches convergence in less than half a minute. Lastly, we remark that the time needed for convergence is sensitive to the different initialization provided to the algorithm, explaining the high variance that characterized the computational costs.

\subsection{Comparison with MCMC} \label{sec:comp_with_MCMC}
We now compare the variational and the MCMC approaches for approximating the Brand posterior. The MCMC algorithm we consider is the modified slice-sampler introduced in \cite{denti2021two}. We compare the estimating approaches leveraging on the same DGP highlighted in the previous section and slightly modified accordingly.
In detail, we consider five spherical bivariate Gaussian to generate the classes, out of which three are deemed as novelties: the basic details of the resulting DGP can be found in the second block of Table \ref{tab:simuall} (SS2).
We consider 50 different scenarios resulting from the interactions of the levels of following three attributes:

\begin{itemize}
    \item \emph{Simple vs. complex scenarios}: we set the variance of all the mixture components to either $\sigma_S^2=0.2$ or $\sigma_C^2=0.75$ (the only exception being $\sigma_{C_5^*}^2=0.375$). The former value implies clear separation among the elements in the simple scenario. In contrast, the latter variance defines the complex case, where we induce some overlap that may hinder the classification. A descriptive plot is displayed in Section S3 of the Supplementary Material;
    
    \item \emph{Sample size}: we modify the default sample size $n = 1000$ by considering different multiplicative factors in $q\in \{0.5, 1, 2.5,5,10\}$, thus obtaining datasets ranging from $500$ to $10000$ observations.
    
    \item \emph{Data dimensionality}: we augment the dimensionality $p$ of the problem by considering  $p\in \{2, 3, 5, 7, 10\}$. The dimensionality augmentation is carried out as described in Section \ref{simu::scaling}.
\end{itemize}

\begin{figure}[t!]
    \centering
    \includegraphics[width = \linewidth]{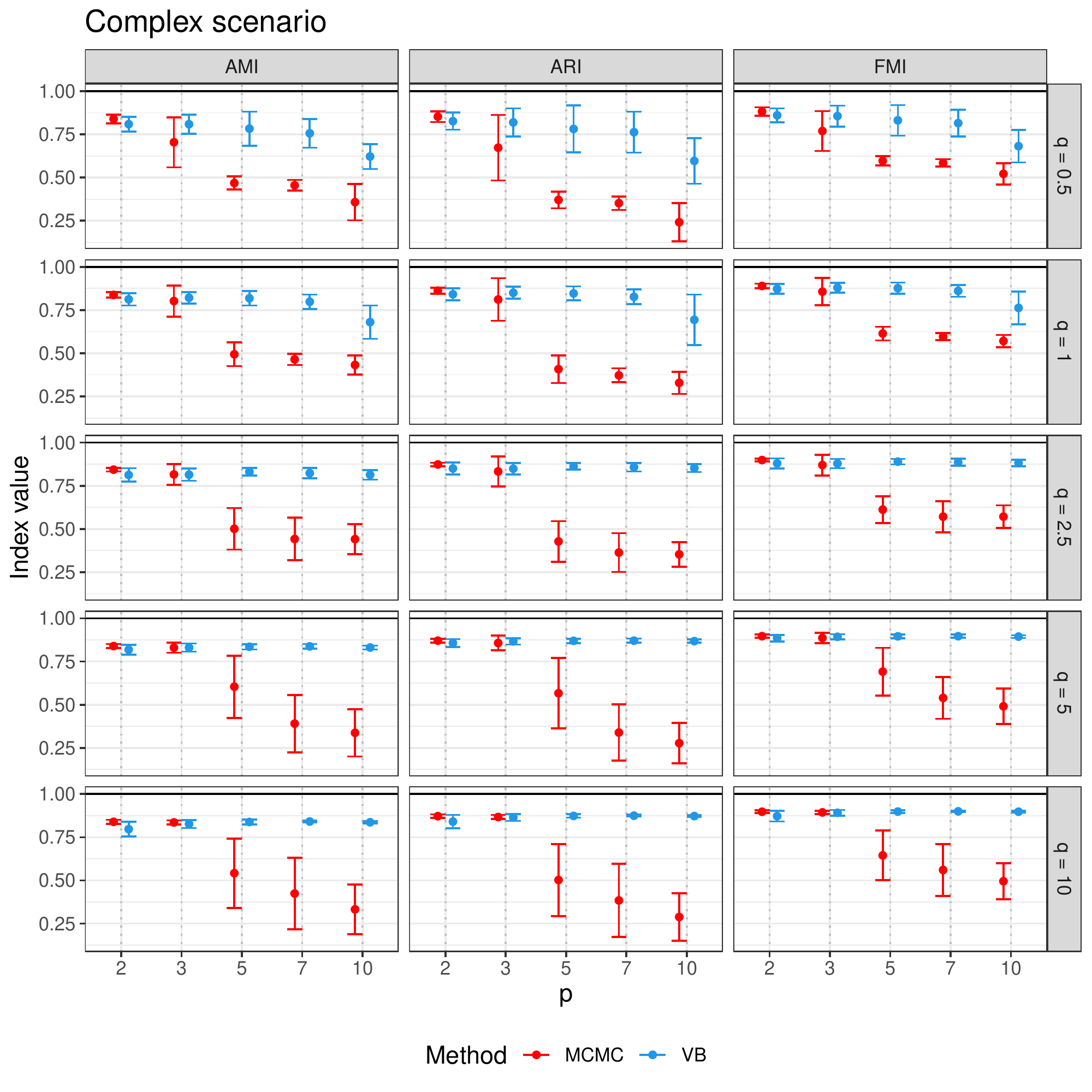}
    \caption{Performance metrics (MCMC in red, variational Bayes - VB -  in blue) stratified by number of variables $p$ and sample size scaling factor $q$ under the complex scenario. The dots represent the averages obtained over 50 replicates of the simulated experiment, while the vertical bars display the associated standard errors.}
    \label{fig:vb_vs_mcmc_ind}
\end{figure}

\begin{figure}[ht]
    \centering
    \includegraphics[width = \linewidth]{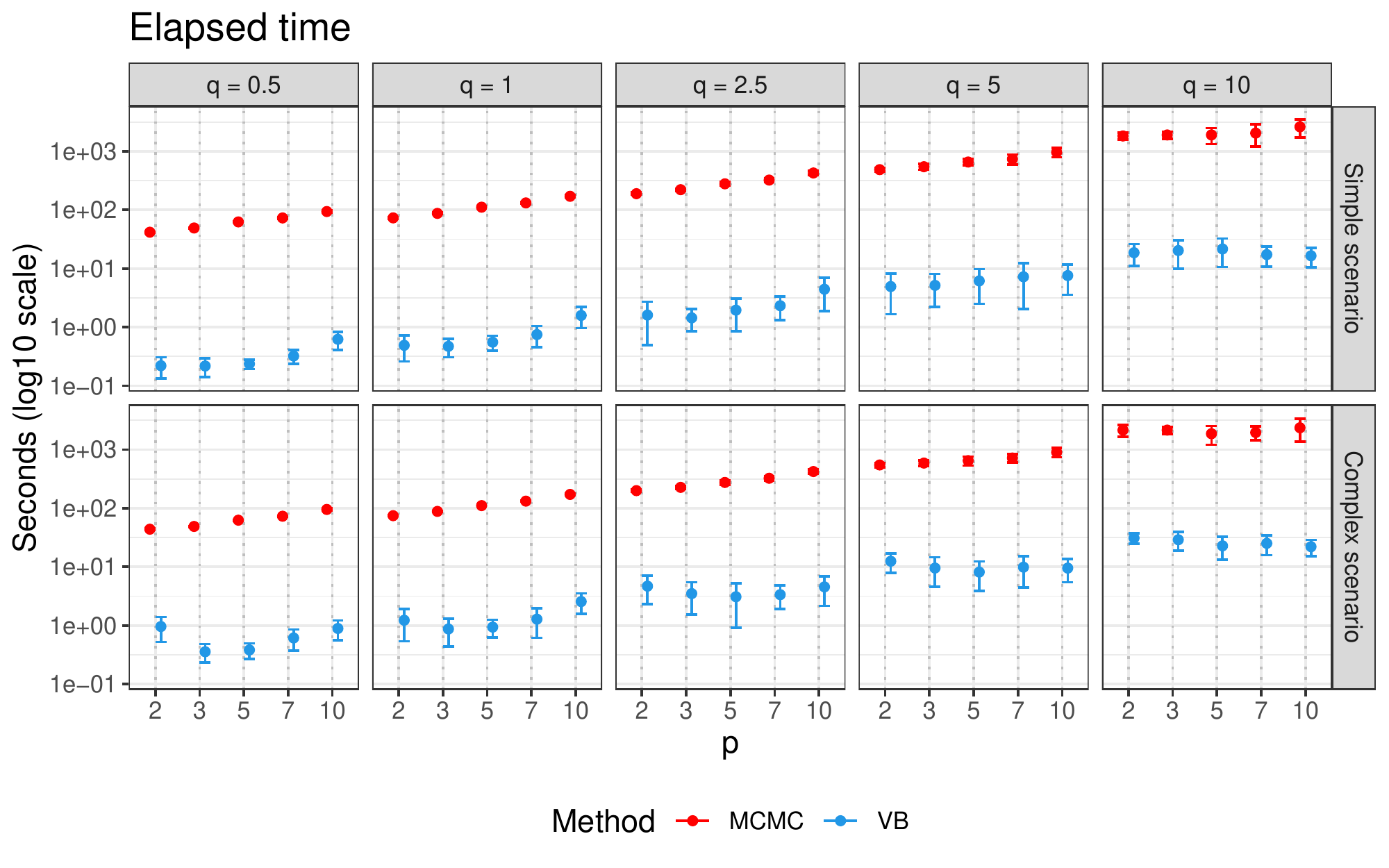}
    \caption{Computational time in seconds (MCMC in red, variational Bayes - VB -  in blue) grouped by number of variables $p$, sample size scaling factor $q$, and type of scenario. The dots represent the averages obtained over 50 replicates of the simulated experiment, while the vertical bars display the associated standard errors.}
    \label{fig:vb_vs_mcmc_time}
\end{figure}

We assess the classification performances with the same metrics previously introduced. For each of the 50 simulated scenarios, we perform 50 Monte Carlo replicates to assess the variation in the performances. A summary of the classification results under the complex scenario are reported in Figure \ref{fig:vb_vs_mcmc_ind}. The panels show that the slice sampler always outperforms the VB implementation in the two-dimensional case. However, as the dimensionality of the dataset increases the MCMC performance rapidly drops, while \texttt{VBrand} always obtains good clustering recovery,  irrespective of the data dimensionality and the sample size. Similar results are obtained under the simple scenario, for which a summarizing plot can be found in Section S3 of the Supplementary Material.
Figure \ref{fig:vb_vs_mcmc_time} compares the algorithms in terms of computation time. As expected, \texttt{VBrand} provides results in just a fraction of the time required by the MCMC approach, being approximately two orders of magnitude faster.

These results cast light not only on the apparent gain in computational speed when using the variational approach, which is expected  by any means, but also on the superior recovering of the underlying partition in the test set in more complex scenarios. 

\subsection{Sensitivity Analysis}
\begin{figure}[h!]
    \centering
    \includegraphics[width=\linewidth]{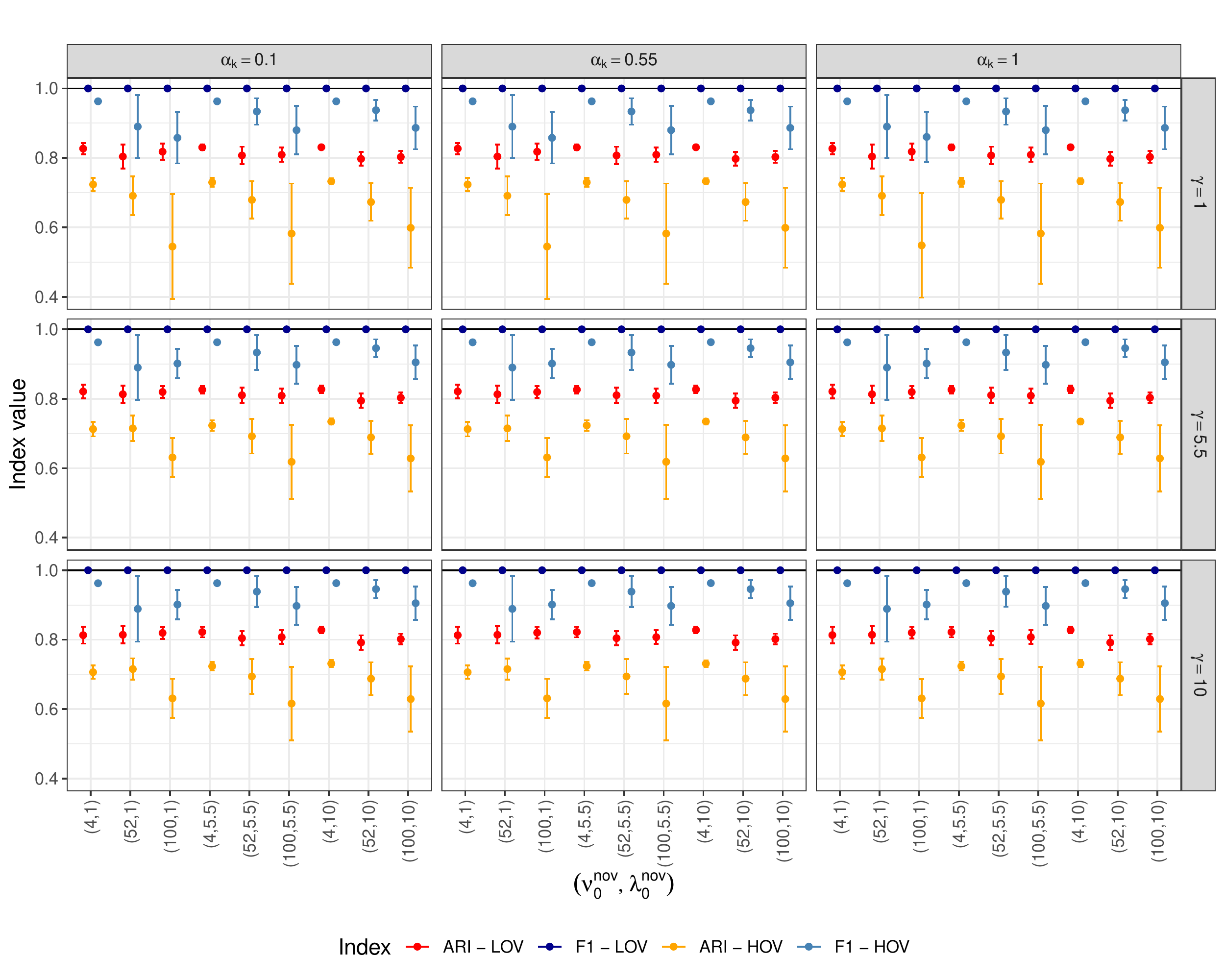}
    \caption{Classification metrics obtained over 50 replicates for 81 combinations of the hyperparameters $\boldsymbol{\alpha}$, $\gamma$, $\lambda^{nov}_0$, and $\nu^{nov}_0$ under the low overlap (LOV) and high overlap (HOV) cases. The dots represent the averages obtained over 50 replicates of the simulated experiment, while the vertical bars display the associated standard errors.}
    \label{fig:SA}
\end{figure}

Finally, we investigate the sensitivity of the model classification to different hyperprior specifications. Two pairs of crucial hyperparameters may considerably affect the clustering results. The first two are $\boldsymbol{\alpha}$ and $\gamma$, which drive the parametric and nonparametric mixture weights, respectively. The second pair is given by the $\mathcal{NIW}$ precision $\lambda^{nov}_0$ and degrees of freedom $\nu^{nov}_0$ of the novelty components.
To assess their impact, we devise a sensitivity analysis considering each possible combination of the following hyperparameters: $\boldsymbol{\alpha} \in \{0.1,0.55,1\}$, $\gamma\in\{1,5.5,10\}$,  $\lambda^{nov}_0\in\{1,5,10\}$, and $\nu^{nov}_0\in \{4,52,100\}$, thus defining 81 scenarios.
We fit \texttt{VBrand} to a dataset composed of five Normals in a bivariate domain, considering a fixed sample size for both training and test sets. We chose a small sample size for the training set to limit the informativeness of the robust estimation procedure. 
Moreover, for each combination of the hyperparameters, we consider both low and high values for the variances of the mixing components, obtaining scenarios with low overlap (LOV) and high overlap (HOV), respectively. Additional details about the data-generating process can be found in the third block of Table \ref{tab:simuall} (SS3). For this experiment, we compare the retrieved partitions in terms of ARI, as done in Sections \ref{simu::scaling} and \ref{sec:comp_with_MCMC}, and by monitoring the $F1$ score, i.e., the harmonic mean of precision and recall. The results are reported in Figure \ref{fig:SA}. We immediately notice by inspecting the panels that for the LOV case the method performs well regardless of the combination of hyper-parameters chosen for the prior specification. In the HOV scenario the recovery of the underlying true data partition is less effective, as it is nevertheless expected. In particular, it seems that setting a high value for the degrees of freedom $\nu^{nov}_0$ produces a slight drop in the ARI metric. This behavior is due to the extra flexibility allowed to the novelty component, by which some of the units belonging to the known groups are incorrectly captured by the novelty term.

In summary, our proposal increases the scalability of the approach introduced in \cite{denti2021two}, allowing for the original novelty detector to be successfully applied to complex scenarios in a more timely manner, without being it affected by the hyper-parameters specification.

\section{Application to novel soil type detection}
\label{sec:appl}

For our application, we consider the \texttt{Statlog (Landsat Satellite)} Data Set, publicly obtained from the UCI machine-learning repository\footnote{https://archive.ics.uci.edu/ml/datasets/Statlog+\%28Landsat+Satellite\%29}. It consists of a collection of observations from satellite images of different soils. Each image contains four spectral measurements over a 3x3 grid, recorded to classify the soil type captured in the picture. There are a total of six different soil types recorded: Red Soil (RS), Grey Soil (GS), Damp Grey Soil (DGS), Very Damp Grey Soil (VDGS), Cotton Crop (CC), and Soil with Vegetation Stubble (SVS).
We frame the original classification problem in a novelty detection task by removing the images of CC and SVS from the training set, leaving these groups in the test set to be detected as novelties. 

Even when performing a simple classification, a method that can account for the possible presence of previously unseen scenarios can be of paramount utility in many fields. For example, new plants \citep{Christenhusz2016}, animals \citep{Camilo2011}, and viruses \citep{Woolhouse2012} are progressively discovered every year. Likewise, related to our application, landscapes present novelties at increasing rates \citep{Finsinger2017}.  
Moreover, a scalable model that can discern and separate outlying observations is necessary when dealing with real-world data, allowing the results to be robust to outliers or otherwise irregular observations. Once these observations are flagged, they can be the objective of future investigations. Thus, our novelty detection application to the \texttt{Statlog} dataset is a nontrivial example that may inspire future applications of our method. 

The original data are already split into training and test sets. After removing the CC and SVS classes from the training set, we obtain a training set of 3486 observations. The test set instead contains 2000 instances. Each observation includes the four spectral values recorded over the 9-pixel grids. Therefore, we will model these data with a semiparametric mixture of 36-dimensional multivariate Normals.
Given the large dimensionality of the dataset, the application of the MCMC estimation approach is problematic in terms of both required memory and computational time. Indeeed, estimating the model via slice sampler becomes unfeasible for most laptop computers. Moreover, we recall that the MCMC approach showed some numerical instabilities in our simulation studies when applied to large dimensional datasets. 

We apply \texttt{VBrand} adopting a mixture with full covariance matrices to capture the potential dependence across the different pixels. Being primarily interested in clustering, we first rescale the data to avoid numerical problems. In detail, we divided all the values in both training and test by 4.5 to reduce the dispersion. Indeed, before the correction, the variabilities within groups ranged from 25.40 to 228.04. After, the within-group variability ranges from 1.25 to 11.26, significantly improving the stability of the algorithm.

Since the variational techniques are likely to find a locally optimal solution, we run the CAVI algorithm 200 times adopting different initializations. For each run, we obtain different random starting point as follows:
\begin{itemize}
    \item we set the centers for the novelty NIWs equal to the centers returned by a $k$-means algorithm performed over the whole test set, with $k$ being equal to the chosen truncation;
       \item the Dirichlet parameters, $\boldsymbol{\nu}^{nov}$ and $\boldsymbol{\lambda}^{nov}$ are randomly selected. In particular, we sample the Dirichlet parameters from $\left(0.1, 1\right)$ and the $\mathcal{NIW}$ parameters from $\left(1,10\right)$ through Latin Hypercube sampling;
\end{itemize}
The other variational hyperparameters are assumed fixed, equal to the corresponding prior hyperparameters. In Section S4 of the Supplementary Material, we report a detailed list of the hyperparameter specifications that we adopted for this analysis.

\begin{figure}[t!]
    \centering
    \includegraphics[width= \linewidth]{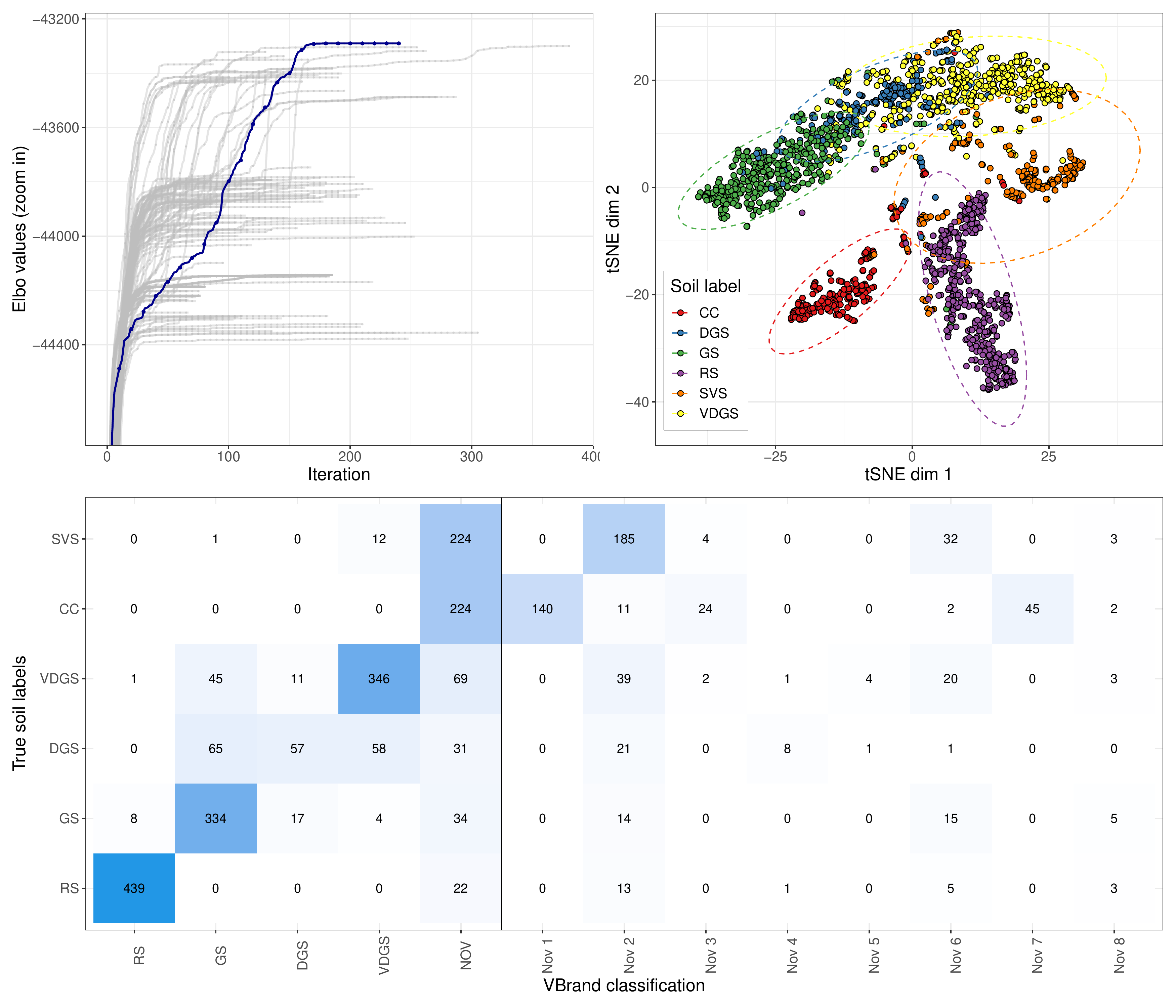}
    \caption{Top-left panel: collection of ELBO trajectories obtained via CAVI updates starting from 200 different random intializations; the trajectory providing the highest ELBO is highlighted in blue (the $y$ axis is truncated for improved visualization). Top-right panel: projection of the test dataset onto a two-dimensional space via the \texttt{tSNE} algorithm. Bottom panel: heatmap of the resulting confusion matrix.}
    \label{fig:t-results}
\end{figure}

As a final result, we select the run with the highest ELBO value, being the one whose variational distribution has the lowest KL divergence from the true posterior. The top-left panel of Figure \ref{fig:t-results} shows the ELBO trends for all the run we performed.
The bottom panel of Figure \ref{fig:t-results} reports the resulting confusion matrix: we observe that the algorithm successfully detected both novelties, achieving a satisfactory classification performance of the previously observed soil types. However, the model struggles with classifying the DGS instances, often mistaken for GS or VDGS. Such difficulty is explained by the overlap between these groups, as shown by the visualization of the test set obtained via the \texttt{tSNE} projection \citep{Hinton2008}, reported in the top-right panel of Figure \ref{fig:t-results}: from the plot, we see that it is not straightforward to establish clear boundaries between GS, DGS, and VDGS soil types.

Overall, \texttt{VBrand} captures the main traits of the data and flags some observations as outliers (e.g., Novelty clusters 4 and 5), which may warrant further investigation.
All in all, our variational approach provides a good clustering solution in a few seconds and it is fast enough to allow for a brute-force search for a better, albeit only locally optimal, solution employing multiple initializations.

\section{Discussion and conclusions}
\label{sec:disc}

In this paper we introduced \texttt{VBrand}, a variational Bayes algorithm for novelty detection, to classify instances of a test set that may conceal classes not observed in a training set. We showed how \texttt{VBrand} outperforms the previously proposed slice sampler implementation in terms of both computational time and robustness of the estimates. The application to soil data provides an example of the versatility of our method in a context where the MCMC algorithm fails because of the large dimensionality of the problem.\\
Our results pave the way for many possible extensions. First, given recent developments in Bayesian nonparametric literature, the variational algorithm can be enriched by adding a hyperprior distribution for the concentration parameter of the novelty DP \citep{Ascolani2022}. While in practice \texttt{VBrand} already obtains very good classification performance, this addition would lead to a consistent model for the number of true clusters. Second, we can consider different likelihood specifications and develop variational inference novelty detectors for, but not limited to, functional or graphical data. Third, at the expense of efficiency, we can explore more complex specifications for the variational distributions, as in structured variational inference. Albeit potentially slower, this choice would lead to an algorithm that could better capture the complex structure of the posterior distribution we are targeting. Finally, we can resort to stochastic variational inference algorithms \citep{Hoffman2003}, to scale up \texttt{VBrand}'s applicability to massive datasets that could benefit from novelty detection techniques.

\section{Acknowledgments}
The authors are grateful to Fabio Lavezzo for his help in the early-stage research that led to the development of this article.

\bibliographystyle{plainnat}
\bibliography{main_scrartcl}

\newpage
\setcounter{section}{0}
\renewcommand*{\thesection}{S\arabic{section}}

\section*{Supplementary Material}

This Supplementary Material contains the formulas to implement the full conditionals of the Brand Gibbs sampler in Section \ref{suppsec::gibbs}. In Section \ref{suppsec::cavi_elbo}, we report additional details that we used to derive the updating rules of the CAVI algorithm and the value of the ELBO. Section \ref{suppsec::addfig} contains additional figures regarding the simulation studies presented in the main paper.
Finally, we report the hyperparameter specifications we adopted for the application to the \texttt{Statlog} dataset in Section \ref{suppsec::soil}.

\section{Full conditionals for Brand with multivariate Gaussian components}
\label{suppsec::gibbs}
Let $X\mid \cdots$ denote the full conditional distribution for the random variable $X$. Then, the full conditionals for the parameters of the model introduced in \cite{denti2021two} are as follows:

\begin{itemize}
    \item The memberships labels are updated according to
    \begin{equation*}
\mathbb{P}\left[\xi_m = k \mid \cdots\right] = \begin{cases}
       \pi_k \cdot \mathcal{N}(\boldsymbol{y}_m\mid\boldsymbol\Theta_k^{obs})\cdot\mathbbm{1}_{\xi_m = k} &\quad\; k=1,\;...\;,J,\\
      \pi_0\cdot v_k \left(\prod_{l=1}^{k-J-1}(1-v_l)\right)
     \cdot \mathcal{N}(\boldsymbol{y}_m\mid\boldsymbol\Theta_k^{nov})\cdot\mathbbm{1}_{\xi_m = k}, &\quad  k\geq J+1.\\
    \end{cases} 
\end{equation*}

\item The weights of the parametric mixtures and the stick-breaking variables are sampled from 
\begin{equation*}
\boldsymbol\pi\mid \cdots \overset{}{\sim} Dirichlet\left(\{\alpha_k+n_k\}_{k=0}^J\right),
\quad \quad 
v_k\mid \cdots \overset{}{\sim} Beta\left( n_{J+k}+1,\gamma+\sum_{l=k+1}^{\infty} n_{J+l}\right),
\end{equation*}

\item The mixture locations are updated according to
$
\boldsymbol{\Theta}_k^{q}\mid\cdots \overset{}{\sim} \mathcal\mathcal{NIW}(\boldsymbol{\mu}_n,\lambda_n,\nu_n,\boldsymbol{\Psi}_n)$, with $q\in\{obs, nov\}$, where
\begin{equation*}\begin{aligned}
& n_k = \begin{cases}
      \#[\xi_m=k] & \text{if $k$ $\geq$ 1 },\\
      \#[\xi_m>J] & \text{if $k$ = 0 },\\
    \end{cases} \quad \quad 
 \boldsymbol\mu_n = \frac{\lambda_0\cdot\boldsymbol{\mu}_0 + \sum_{m=1}^{M}\boldsymbol{y}_m\cdot\mathbbm{1}_{\xi_m=k}}{\lambda_0+n_k},\\
& \lambda_n=\lambda_0+n_k,\quad \quad  \nu_n=\nu_0+n_k,\\
& \boldsymbol\Psi_n=\boldsymbol\Psi_0+ \boldsymbol{S} + \frac{\lambda_0 \cdot n_k}{\lambda_0+n_k}\left(\frac{\sum_{m=1}^{M} \boldsymbol{y}_m\cdot\mathbbm{1}_{\xi_m=k}}{n_k} - \boldsymbol{\mu}_0\right)\left(\frac{\sum_{m=1}^{M} \boldsymbol{y}_m\cdot\mathbbm{1}_{\xi_m=k}}{n_k} - \boldsymbol{\mu}_0\right)^T,\\
& \boldsymbol{S}=\sum_{m=1}^{M}(\boldsymbol{y}_m-\overline{\boldsymbol{y}}^k)\cdot(\boldsymbol{y}_m-\overline{\boldsymbol{y}}^k)^T\cdot\mathbbm{1}_{\xi_m=k},
\quad \quad  \overline{\boldsymbol{y}}_k = \frac{\sum_{m=1}^M\boldsymbol{y}_m\cdot\mathbbm{1}_{\xi_m=k}}{\sum_{m=1}^M\mathbbm{1}_{\xi_m=k}}.\\
\end{aligned}\end{equation*}

Here, ($\boldsymbol{\mu}_0$, $\lambda_0$, $\nu_0$, $\boldsymbol\Psi_0$) indicate the hyperparameters of the relative ${NIW}$ distribution, according to which atom $\boldsymbol{\Theta}_k^{q}$ is being considered.\\

\end{itemize}

\section{Additional details regarding the VBrand algorithm}
\label{suppsec::cavi_elbo}
\subsection{Quantities useful to devise the CAVI updating rules}

Here, we report some additional formulas to clarify the derivation of the CAVI algorithm for VBrand. In our implementation, we used the following well-known expected values:

\begin{itemize}
    \item $
    \mathbb{E}[\log{\pi_k }] = \boldsymbol{\psi}(\eta_k)-\boldsymbol{\psi}(\overline{\eta}),$ with $\overline{\eta} = \sum_k \eta_k$

    \item $
    \mathbb{E}[\log{v_{k-J}}] = \boldsymbol{\psi}(a_{k-J})-\boldsymbol{\psi}(a_{k-J}+b_{k-J})$ and $ \mathbb{E}[\log{(1-v_{l})}] = \boldsymbol{\psi}(b_{l})-\boldsymbol{\psi}(a_{l}+b_{l})$,

\item If $\Sigma^{-1} \sim \textit{Wishart}$, we have that:
\begin{equation*}
\mathbb{E}[\log{\mid\Sigma^{-1}\mid}] = -p\log{2} - \log{\mid\boldsymbol{\Psi}_k^{-1}\mid} - \sum_{l=1}^p\boldsymbol{\psi}\left(\frac{\nu_k-l+1}{2}\right).
\end{equation*}
\end{itemize}
Given the previous results and the updating rules derived in Section \ref{suppsec::gibbs}, we can easily derive the value for the negative entropy of a multivariate Normal distribution for $q\in\{obs,nov\}$:
 \begin{equation*}\begin{split}
\mathbb{E}[\log{\mathcal{N}(\boldsymbol{y}_m\mid\boldsymbol{\Theta}_k^{q})}] &= \frac{1}{2}\left(-p\log{2\pi} + \sum_{l=1}^p\boldsymbol{\psi}\left(\frac{\nu_k-l+1}{2}\right) + p\log{2} + \log{\mid\boldsymbol{\Psi}_k^{-1}\mid} \right) + \\& -\frac{1}{2} \left( \frac{p}{\lambda_k} +
\nu_k(\boldsymbol{y}_m-\boldsymbol{\mu}_k)^T\boldsymbol{\Psi}_k^{-1}(\boldsymbol{y}_m-\boldsymbol{\mu}_k)\right).
\end{split}\end{equation*}

\subsection{Quantities useful to derive the Elbo components}

In the following formulas, $Tr(\cdot)$ denotes the trace of a matrix. Moreover, the expression of the function $B(\cdot,\cdot)$ can be found in \cite{Bishop2006}, Appendix B, p.693.

\begin{equation*}\begin{split}
 \mathbb{E}[\log( \mathcal{NIW}(\boldsymbol{\Theta}_k^{obs}\mid\boldsymbol{\varrho}_k
 ))] =& \frac{1}{2}\big(p\log{\frac{\lambda_k^{obs}}{2\pi}} + \sum_{i=1}^p\psi(\frac{u_k^{obs}-i+1}{2}) + \log{\mid\boldsymbol{S}_k^{obs^{-1}}\mid} +\\ 
 &- p\frac{\lambda_k^{obs}}{l_k^{obs}} - \lambda_k^{obs}u_k^{obs}(\boldsymbol{m}_k^{obs}-\boldsymbol{\mu}_k^{obs})^T\boldsymbol{S}_k^{obs^{-1}}(\boldsymbol{m}_k^{obs}-\boldsymbol{\mu}_k^{obs})\big) +\\
 &+\log{B(\boldsymbol{\Psi}_k^{obs^{-1}},\nu_k^{obs})} - \frac{1}{2}u_k^{obs} Tr(\boldsymbol{\Psi}_k^{obs}\boldsymbol{S}_k^{obs^{-1}}) + const,
\end{split}\end{equation*}

\begin{equation*}\begin{split}
 \mathbb{E}[\log( \mathcal{NIW}(\boldsymbol{\Theta}_k^{nov}\mid\boldsymbol{\varrho}_0
 ))] =& \frac{1}{2}\big(p\log{\frac{\lambda_0^{nov}}{2\pi}} + \sum_{i=1}^p\psi(\frac{u_k^{nov}-i+1}{2}) + \log{\mid\boldsymbol{S}_k^{nov^{-1}}\mid} +\\ 
 &- p\frac{\lambda_0^{nov}}{l_k^{nov}} - \lambda_0^{nov}u_k^{nov}(\boldsymbol{m}_k^{nov}-\boldsymbol{\mu}_0^{nov})^T\boldsymbol{S}_k^{nov^{-1}}(\boldsymbol{m}_k^{nov}-\boldsymbol{\mu}_0^{nov})\big) +\\
 &+\log{B(\boldsymbol{\Psi}_0^{nov^{-1}},\nu_0^{nov})} - \frac{1}{2}u_0^{nov} Tr(\boldsymbol{\Psi}_0^{nov}\boldsymbol{S}_k^{nov^{-1}}) + const,
\end{split}\end{equation*}

\begin{equation*}\begin{aligned}
 \mathbb{E}[\log( \mathcal{NIW}(\boldsymbol{\Theta}_k^{obs}\mid\boldsymbol{\rho}_k^{obs}))]\ = \frac{u_k^{obs}}{2}\log{\mid\boldsymbol{S}_k^{obs}\mid} - \frac{u_k^{obs} p}{2}\log{2} - \log\Gamma_p\left(\frac{u_k^{obs}}{2}\right) + \frac{p}{2}\log{l_k^{obs}} + \\
 + \frac{u_k^{obs} + p + 2}{2}  \left(\log{\mid\boldsymbol{S}_k^{obs^{-1}}\mid} + \sum_{i=1}^p\boldsymbol{\psi}\left(\frac{u_k^{obs}-i+1}{2}\right)\right) - \frac{u_k^{obs}}{2} p,
 \end{aligned}\end{equation*}

\begin{equation*}\begin{aligned}
 \mathbb{E}[\log( \mathcal{NIW}(\boldsymbol{\Theta}_k^{nov}\mid\boldsymbol{\rho}_k^{nov}))]\ = \frac{u_k^{nov}}{2}\log{\mid\boldsymbol{S}_k^{nov}\mid} - \frac{u_k^{nov} p}{2}\log{2} - \log\Gamma_p\left(\frac{u_k^{nov}}{2}\right) + \frac{p}{2}\log{l_k^{nov}} + \\
 + \frac{u_k^{nov} + p + 2}{2}  \left(\log{\mid\boldsymbol{S}_k^{nov^{-1}}\mid} + \sum_{i=1}^p\boldsymbol{\psi}\left(\frac{u_k^{nov}-i+1}{2}\right)\right) - \frac{u_k^{nov}}{2} p.
\end{aligned}\end{equation*}

\newpage
\section{Additional figures}
\label{suppsec::addfig}

\begin{figure}[!ht]
	\centering
\includegraphics[width =.8\linewidth]{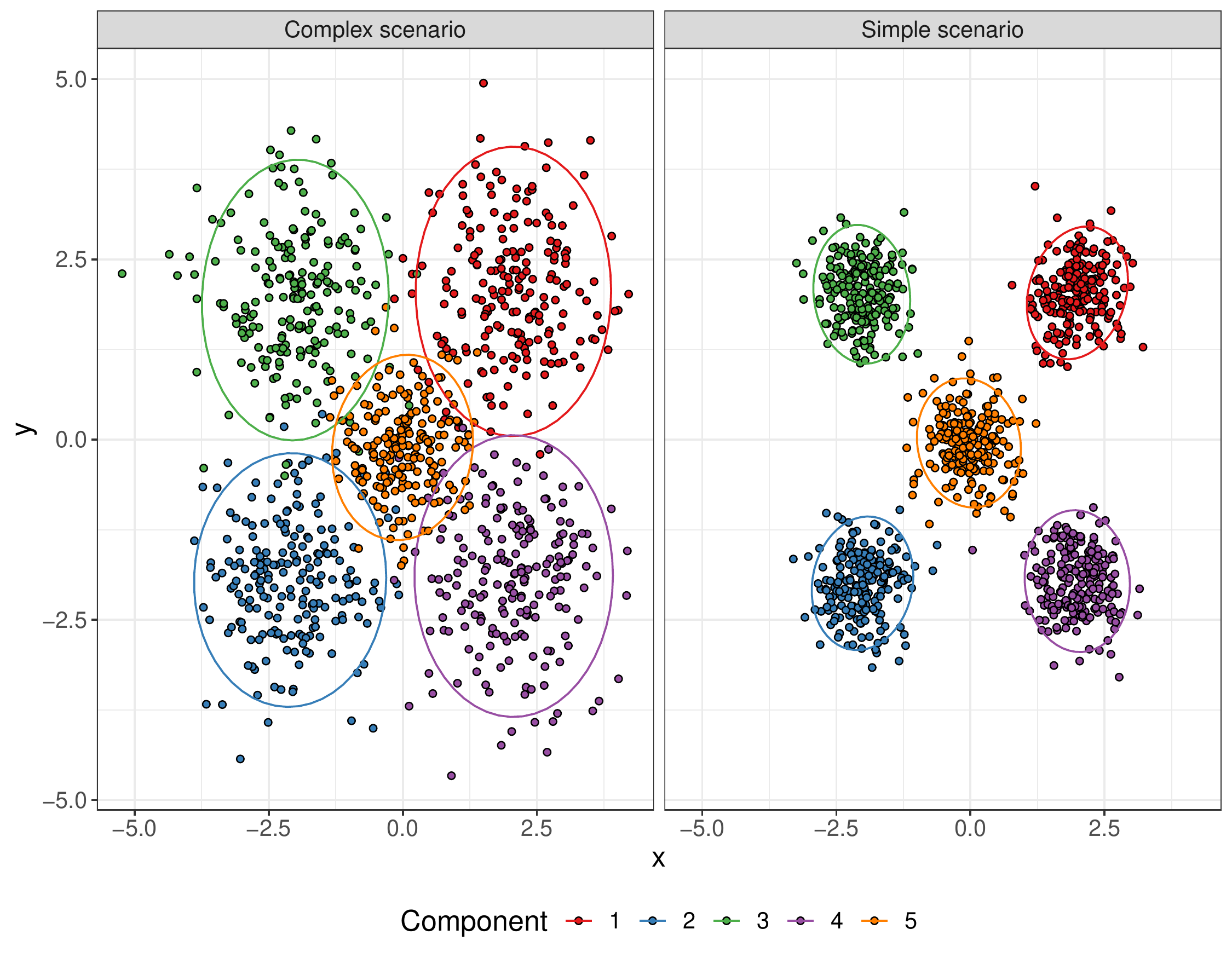}
\caption{Example of the first two dimensions of the generated data under the simulation study described in Section 4.2 of the main paper.}
\label{fig:supp_simple_vs_complex}
\end{figure}
\FloatBarrier
\begin{figure}[!ht]
	\centering
\includegraphics[width = .8\linewidth]{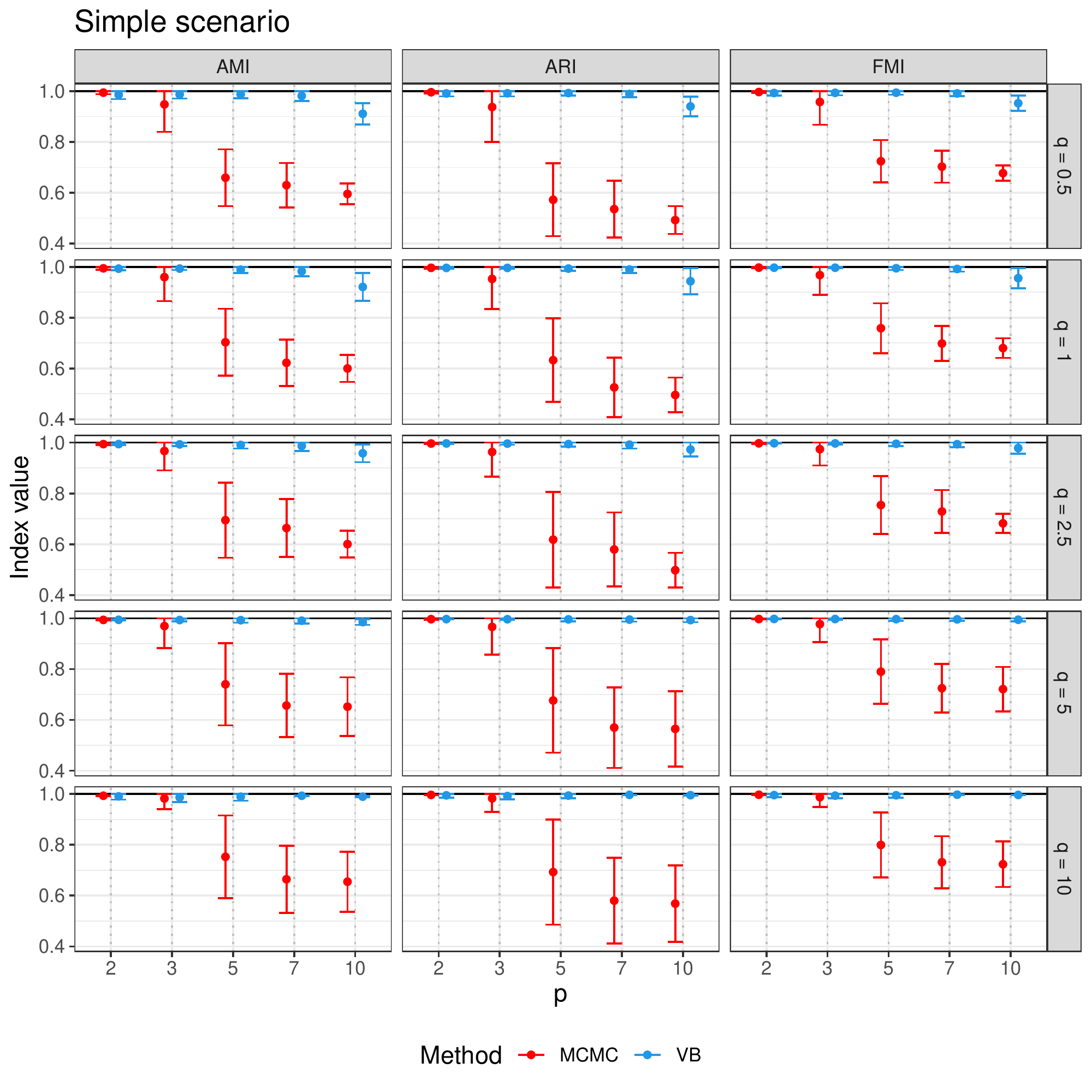}
\caption{Results of the simulation study described in Section 4.2 of the main paper under the simple scenario.}
\label{fig:supp_simple_res}
\end{figure}
\FloatBarrier

\clearpage
\section{Hyperparameter initializations for the novel soil detection analysis}

\label{suppsec::soil}
We report a detailed list of the hyperparameters initialization of the VBrand model used for the novelty detection of soil type in Section 5 of the main paper.

\begin{enumerate}
    \item The truncation is set at $T=10$. \item The concentration parameter of the stick-breaking process is set to $\gamma=5$. 
    \item The hyperparameters of the Dirichlet distribution are all set to $\alpha_j = 0.1\;\;\forall j$, inducing the same probability of being in each known component and of being a novelty.
    \item The $\mathcal{NIW}$ prior for the novelty locations is parameterized as follows: $\boldsymbol{\mu}_0^{nov}$ is set equal to the grand mean of the training set, to allow an hyperprior specification that does not rely on the test set; the precision is given as 
    ${\lambda}_0^{nov} =  0.1$,
    the degrees of freedom ${\nu}_0^{nov}= p+2$, where $p$ is the dimensionality of the dataset. This is the minimum integer value that is allowed to have so that the $\mathcal{IW}$ distribution has a finite mean. Finally, the variance is set to be $(p+1)$ times the training sample covariance matrix. This choice implies that the expected variance sampled from the $\mathcal{NIW}$ is an inflated version of the overall sample covariance matrix of the training set.
    \item For the known components, the mean vector, and the variance-covariance matrices are estimated via the MRCD procedure. To put high mass on these estimates, for all the known groups, we set $\lambda_k^{obs}= 200$ and $\nu^{obs}_k =  p+1+200$. 
\end{enumerate}

\end{document}